# Superconductivity of high-entropy-alloy-type transition-metal zirconide (Fe,Co,Ni,Cu,Ga)Zr$_2$

Md. Riad Kasem[1], Hiroto Arima[1], Yoichi Ikeda[2], Aichi Yamashita[1] and Yoshikazu Mizuguchi[1]

[1] Department of Physics, Tokyo Metropolitan University, Hachioji, Japan
[2] Institute for Materials Research, Tohoku University, Sendai, Japan

E-mail: mizugu@tmu.ac.jp

**Abstract**

We synthesized a new high-entropy-alloy-type (HEA-type) superconductor (Fe,Co,Ni,Cu,Ga)Zr$_2$ with a $T_c$ of 2.9 K. The EDX analyses revealed that the actual composition of the transition-metal site ($Tr$-site) is $Tr$ = Fe$_{0.18}$Co$_{0.18}$Ni$_{0.16}$Cu$_{0.25}$Ga$_{0.23}$, which gives the configurational entropy of mixing $\Delta S_{mix}$ = 1.60$R$ for the $Tr$ site. Neutron powder diffraction revealed that the sample has a tetragonal CuAl$_2$-type (space group: #140). The lattice constant of $a$ monotonically decreases with decreasing temperature, but the lattice constant of $c$ does not exhibit a clear shrinkage. Isotropic displacement parameter for both the $Tr$ and Zr sites are large, which is probably caused by the HEA-type $Tr$ site. The small temperature dependences of $U_{iso}$ for both sites also indicate the presence of the local structural disorder in (Fe,Co,Ni,Cu,Ga)Zr$_2$. From electrical resistivity, magnetic susceptibility, and specific heat measurements, bulk superconductivity was confirmed.

## 1. Introduction

Transition-metal zirconides ($Tr$Zr$_2$) are superconductors with a wide range of transition temprature: $T_c$ = 1.6, 5.5–6.0, 11.3, 7.5 K for $Tr$ = Ni, Co, Rh, Ir, respectively [1]. The flexible solution at the $Tr$ site is possible, and $T_c$ is tuned by the $Tr$-site solution [2–4]. In addition, CoZr$_2$ exhibits a large enhancement of $T_c$ under pressures, which suggests that the superconducting properties of $Tr$Zr$_2$ are sensitive to crystal-structure modification [5]. Recently, we reported synthesis and superconductivity of (Fe,Co,Ni,Rh,Ir)Zr$_2$ and (Co,Ni,Cu,Rh,Ir)Zr$_2$ [6,7], in which the $Tr$ site was designed based on the high-entropy-alloy (HEA) concept [8–12]. Due to the solution of five or more $Tr$ elements at the $Tr$ site, local structural disorders are expected, while investigation on local structural disorder has not been addressed in detail. Instead, from specific heat measurements on the superconducting transitions for the HEA-type $Tr$Zr$_2$, local (microscopic) inhomogeneity of the superconducting states were revealed from the broadening of the specific heat jump [13]. Because creation of local structural disorder and/or inhomogeneity of superconducting states would be useful for developing superconductivity application and creating novel superconducting states [14–19], further development of the examples of HEA-type superconductors with a high purity is desired. In this study, we synthesized an HEA-type $Tr$Zr$_2$ superconductor with 3d elements (Fe, Co, Ni, Cu, and Ga) and Zr, while we have investigated HEA-type $Tr$Zr$_2$ superconductors containing 4d and 5d elements [6,7,13]. The purity of the CuAl$_2$-type phase in the examined (Fe,Co,Ni,Cu,Ga)Zr$_2$ was the best among the HEA-type $Tr$Zr$_2$ samples studied so far. Therefore, we could reveal the effects of the HEA-type $Tr$ site on the local disorder using neutron powder diffraction (NPD) at different temperatures. In addition, anomalous temperature evolution on the lattice constant $c$ was revealed as well. Recently, we reported on the negative thermal expansion of the $c$-axis for $CoZr_2$ and alloyed phases [20]. The anomalous $c$-axis evolution in $(Fe,Co,Ni,Cu,Ga)Zr_2$ would be related to the negative thermal expansion of the $c$-axis observed in $CoZr_2$.

## 2. Experimental details

The Polycrystalline samples of $(Fe,Co,Ni,Cu,Ga)Zr_2$ was prepared by arc melting. Powders of Fe (99.9%), Co (99%), Ni (99.9%) and Cu (99.9%) and grains of Ga (99.9999%) were pelletized and melted together with Zr foils (99.2%) in an Ar atmosphere. The arc melting was repeated three times for homogenization. The actual composition of the obtained sample was examined by energy-dispersive X-ray spectroscopy (EDX, SwiftED, Oxford) on a scaning electron microscope (TM3030, Hitachi-hightech). From the obtained $Tr$-site composition, we calculated configurational entropy of mixing ($\Delta S_{mix}$) at the $Tr$ site using the fomula of $\Delta S_{mix} = -R\Sigma_i c_i \ln c_i$, where $c_i$ and $R$ are the atomic fraction of component $i$ and the gas constant, respectively.

The purity and crystal structure were investigated by NPD, and the Rietveld method was used for refining the crystal structure parameters. Low-temperature NPD experiments were performed with a HERMES diffractometer [21] installed at the T1-3 guide port in the JRR-3 of the Japan Atomic Energy Agency, Tokai. A thermal neutron beam was monochromatized to be 2.1972 Å with a vertically-focused Ge (331) monochromator. Typical instrumental parameters were determined by analyzing line positions and line shapes of a standard reference material ($LaB_6$, NIST660c) [22]. Powder samples were sealed in a vanadium cylinder cell with $\phi$6 mm-diameter (thickness 0.1 mm) and ~60mm-length in a $^4$He gas atmosphere. A closed-cycle refrigerator was use to cool the samples and controlled from a base temperature ($T \sim$ 2 K) to room temperature. The obtained XRD and NPD patterns were refined by the Rietveld method using RIETAN-FP [23], and the schematic images of the refined crystal structure were depicted using VESTA [24].

The superconducting properties were investigated by electrical resistivity, magnetic susceptibility, and specific heat experiments. The temperature dependence of electrical resistivity was measured by the DC four-probe method under magnetic fields on a Physical Property Measurement System (PPMS, Quantum Design). Au wire ($\phi$25 μm) and Ag paste were used for fabricating the terminals. The temperature dependence of magnetic susceptibility was measured by a superocnducting intereference device (SQUID) magnetometer on a Magnetic Property Measurement System (MPMS3, Quantum Design) in both zero-field-cooling (ZFC) and field-cooling (FC) modes with an applied field of 10 Oe. The temperature dependence of specific heat was measured by the thermal relaxation method on a PPMS.

## 3. Results and discussion

### 3-1. Structural characterization

The actual $Tr$-site composition estimated from EDX is $Tr$ = $Fe_{0.18}Co_{0.18}Ni_{0.16}Cu_{0.25}Ga_{0.23}$, and the estimated $\Delta S_{mix}$ is $1.60R$ for the examined sample. In this paper, we call the HEA-type $TrZr_2$ sample $(Fe,Co,Ni,Cu,Ga)Zr_2$.

Figures 1(a) and 1(b) show the NPD pattern taken at $T$ = 290 K and 2 K, respectively, and the Rietveld refinement results. The profile could be refined with the tetragonal $CuAl_2$-type (space group: #140) model. To obtain better fitting, a small amount of the impurity phase of $TrZr_3$ (7%) is included in the refinements. The reliability factor for $T$ = 290 K is $R_{wp}$ = 11.7%. By comparing NPD patterns at $T$ = 290 and 2 K, we notice that there is no clear indication of the magnetic ordering in the examined sample.

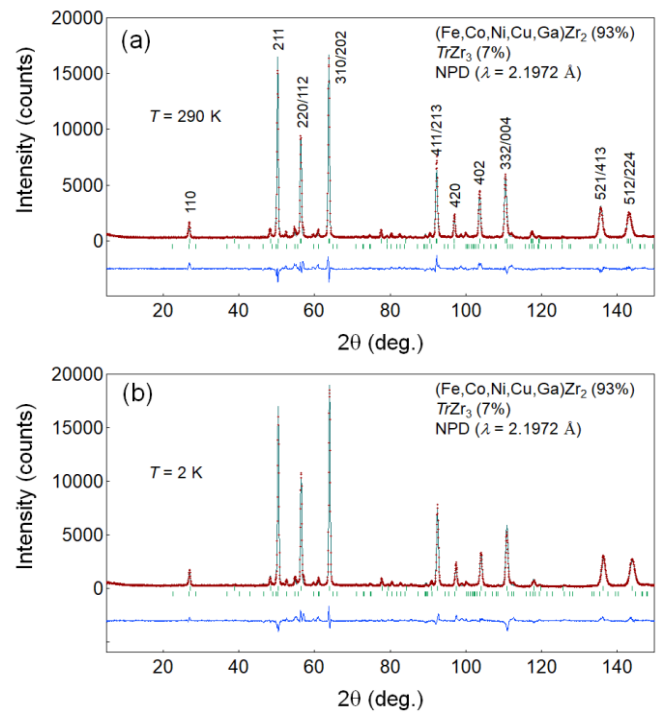

Fig. 1. NPD patterns for $(Fe,Co,Ni,Cu,Ga)Zr_2$ taken at (a) $T$ = 290 K and (b) $T$ = 2 K. The red dots are experimental data. The green and blue curves are the fiting curve and the residual curve, respectively. The ticks indicate the position of the Bragg peaks for the major phase (($Fe,Co,Ni,Cu,Ga)Zr_2$) and the minor phase ($TrZr_3$).

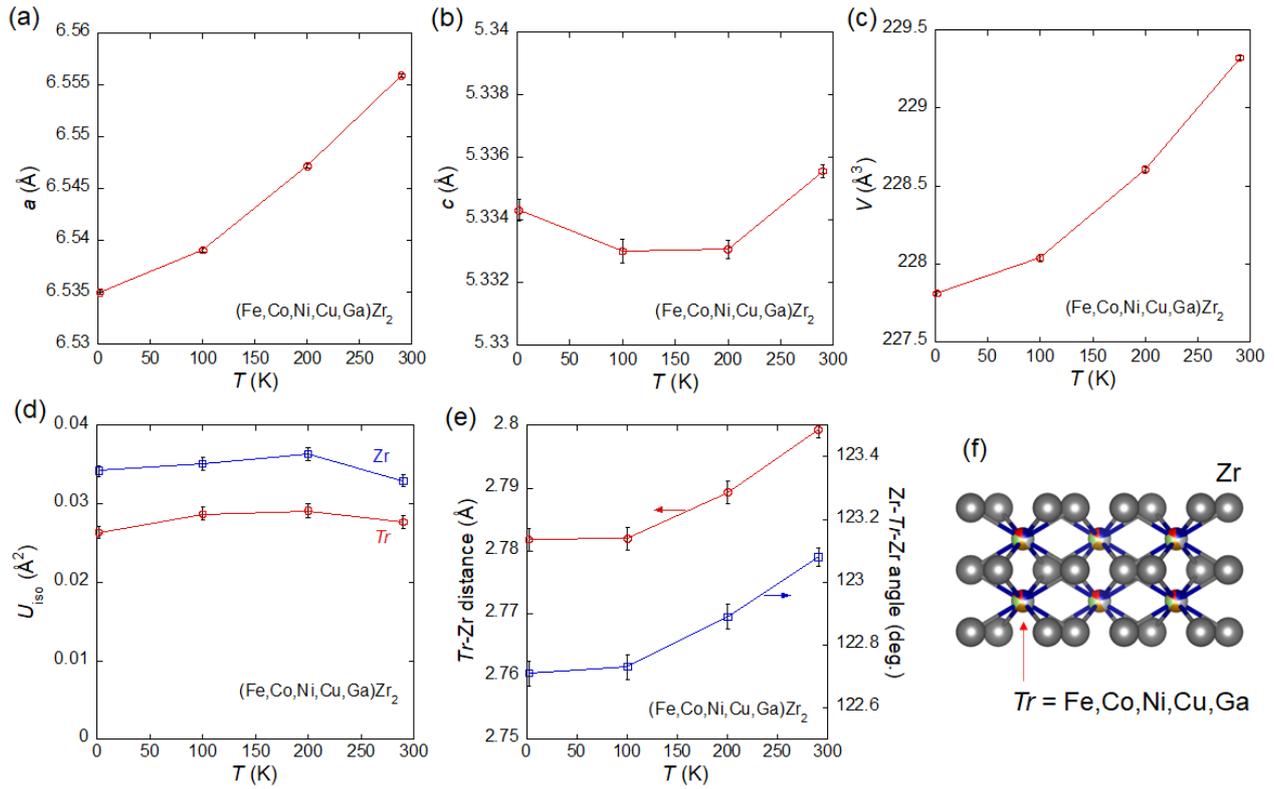

Fig. 2. Crystal-structure parameters for $(Fe,Co,Ni,Cu,Ga)Zr_2$. Temperature dependence of lattice constant (a) $a$, (b) $c$, (c) $V$, and (d) isotropic displacement parameters $U_{iso}$. (e) Temperature dependence of $Tr$-$Zr$ distance and $Zr$-$Tr$-$Zr$ angle. (f) Schematic image of the crystal structure of $(Fe,Co,Ni,Cu,Ga)Zr_2$.

The refined crystal-structure parameters are summarized in Fig. 2. With decreasing temperature the lattice constant $a$ decreases, but $c$ exhibits almost constant value. Because the decrease in $Zr$-$Tr$-$Zr$ angle with decreasing temperature (Fig. 2(e)) is common to $CoZr_2$ [20], the anomalous $c$-axis thermal expansion would be achieved by the flexible structure of the $TrZr_8$ polyhedron. Figure 2(d) shows the temperature dependences of isotropic displacement parameter $U_{iso}$ for the $Tr$ and $Zr$ sites. Noticeably, the $U_{iso}$s do not remarkably change with decreasing temperature for the present sample. To investigate the effects of the presence of the HEA-type $Tr$ site on the temperature dependence of $U_{iso}$ in $TrZr_2$, the data for $CoZr_2$ ($\Delta S_{mix} = 0$) are plotted together in Fig. 3. The NPD results for $CoZr_2$ were reported in Ref. 20. The $U_{iso}$ for Co at $T = 50$ K was too small and could not be refined; hence, the $U_{iso}$ for Co at $T = 50$ K was fixed to $U_{iso}$ for the $Tr$ site. For $CoZr_2$, $U_{iso}$s clearly decrease with decreasing temperature, which is ordinary temperature dependence of atomic displacements with assumption of atomic vibrations. In the case of materials with anharmonic vibrations (as in caged materials) [25] and/or local disorders [26,27], large $U_{iso}$ with a weak temperature dependence has been observed. The large $U_{iso}$s with a small temperature dependence in $(Fe,Co,Ni,Cu,Ga)Zr_2$ indicate the glassy phonon characteristics, which would be induced by the HEA-type $Tr$ site.

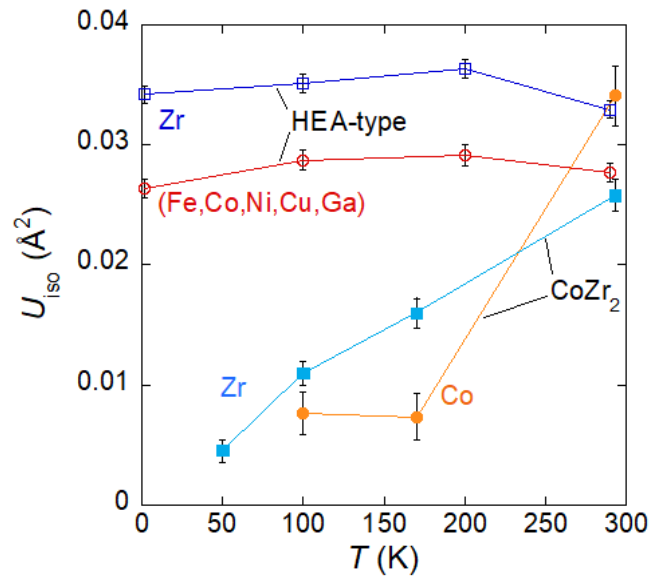

Fig. 3. Temperature dependences of $U_{iso}$ for the $Tr$ and $Zr$ sites in $(Fe,Co,Ni,Cu,Ga)Zr_2$ and $CoZr_2$.

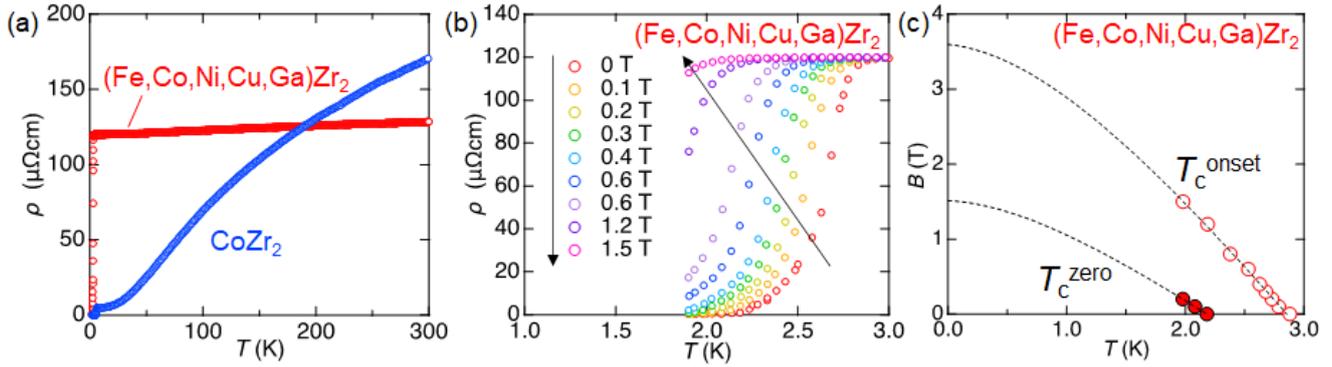

Fig. 4. (a) Temperature dependences of electrical resistivity ($\rho$) for $CoZr_2$ and $(Fe,Co,Ni,Cu,Ga)Zr_2$. (b) Temperature dependences of $\rho$ under magnetic fields for $(Fe,Co,Ni,Cu,Ga)Zr_2$. (c) Temperature dependences of the upper critical field ($B_{c2}$) evaluated from the $T_c^{onset}$ and $T_c^{zero}$ in Fig. 4(b).

### 3-1. Superconducting properties

Figure 4(a) shows the temperature dependences of electrical resistivity ($\rho$) for $CoZr_2$ and $(Fe,Co,Ni,Cu,Ga)Zr_2$. It is clear that the residual resistivity ratio $RRR \sim 1$ for $(Fe,Co,Ni,Cu,Ga)Zr_2$ is larger than that for $CoZr_2$, which is common feature in HEA-type compounds [12] and a clear indication of the local disorder introduced onto the HEA-type $Tr$ site. Figures 4(b) and 4(c) show the $\rho$-$T$ and the upper critical fields ($B_{c2}$) estimated from the $T_c^{onset}$ and $T_c^{zero}$; to estimate $B_{c2}(0)$, we used the Werthamer-Helfand-Hohenberg (WHH) model [28] and the fitting results are displayed in Fig. 4(c). The estimated $B_{c2}$s from $T_c^{zero}$ and $T_c^{onset}$ are 1.5 T and 3.5 T, respectively.

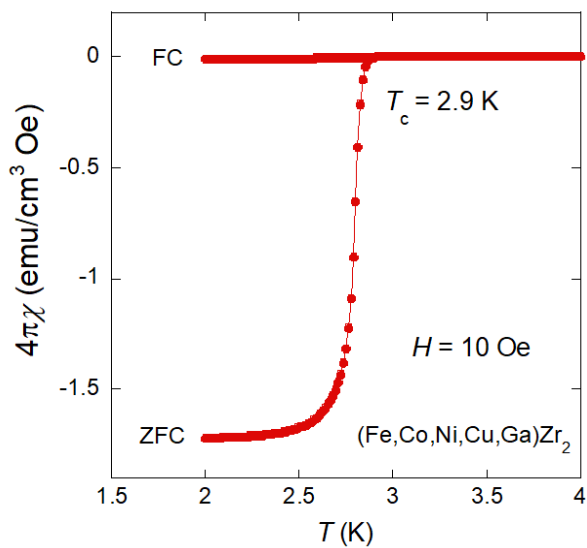

Fig. 5. Temperature dependence of magnetic susceptibility ($4\pi\chi$) for $(Fe,Co,Ni,Cu,Ga)Zr_2$.

Figure 5 shows the temperature dependence of magnetic susceptibility measured after ZFC and FC. A large diamagnetic signals were observed below 2.9 K, which indicates the emergence of bulk superconductivity below $T_c$ = 2.9 K. Figure 6 shows the temperature dependence of electronic specific heat ($C_{el}/T$), where $C_{el}$ was calculated by subtracting phonon contribution ($\beta T^3$) estimated from the low-temperature-limit formula of $C = \gamma T + \beta T^3$. $\gamma$ is electronic-specific-heat coefficient and was estimated as 19.03(4) mJ/K$^2$mol. The jump of $C_{el}$ at $T_c$ was estimated by considering the entropy balance as shown with blue lines in Fig. 6. The estimated $C_{el}/\gamma T_c$ is 1.35 with $T_c$ (specific heat) = 2.78 K, which is almost consistent with the weak-coupling BCS model [29]. From both magnetic susceptibility and specific heat, bulk nature of superconductivity was confirmed.

Here, we briefly discuss about the inhomogeneity of superconducting states seen by specific heat in $TrZr_2$. As reported in Ref. 13, HEA-type $TrZr_2$ containing Rh and/or Ir exhibited clear broadening of the superconducting jump in $C_{el}$ near $T_c$; their $T_c$s were around 5 K. Since no clear phase separation was observed in those HEA-type $TrZr_2$, we concluded that the inhomogeneous superconducting transitions in HEA-type $TrZr_2$ containing Rh and/or Ir are caused by the local disorder introduced by the HEA-type $Tr$ site. However, in the present sample, $(Fe,Co,Ni,Cu,Ga)Zr_2$, the transition is sharp. One of the differences between previous phases containing Rh and/or Ir and the present $(Fe,Co,Ni,Cu,Ga)Zr_2$ is $T_c$. $T_c$ (specific heat) = 2.78 K is lower than 5 K for the samples examined in Ref. 13. Another potential explanation is the difference in electronic structure. By alloying the $Tr$ site in $TrZr_2$, the electronic structure should be affected and the band dispersion would be smeared, as observed in HEAs [30,31]. Since electronic characters of 3d, 4d, and 5d orbitals are different, the influence of the HEA-type $Tr$ site would be smaller in 3d-based $(Fe,Co,Ni,Cu,Ga)Zr_2$ than that in compositions containing 4d of Rh and 5d of Ir. To clarify the electronic structure of HEA-type $TrZr_2$, growth of single crystals and angle-resolved photoemission spectroscopy are desired. We enphasize, however, that the glassy phonon characteristics revealed in Fig. 3 are evidently induced by the introduction of high configurational entropy of mixing at the *Tr* site in *Tr*Zr$_2$. Therefore, the 3d-based (Fe,Co,Ni,Cu,Ga)Zr$_2$ superconductor reported in this paper will be useful to understand the electronic, phonon, and superconducting properties and their relations to local structural disorder and $\Delta S_{mix}$ at the *Tr* site in the *Tr*Zr$_2$ system.

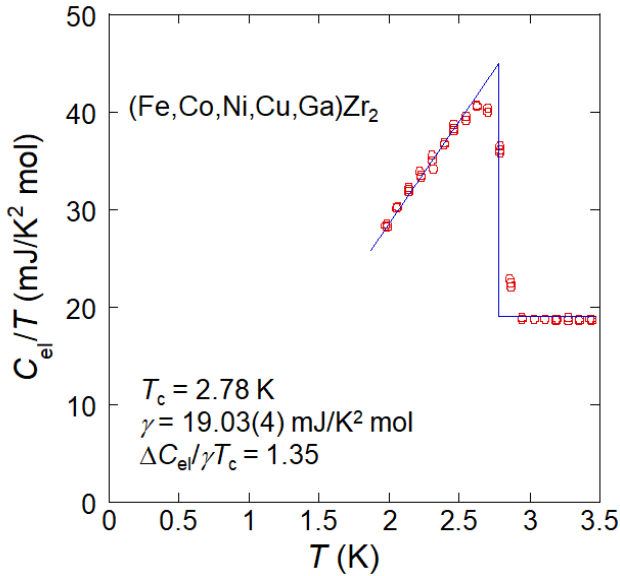

Fig. 6. Temperature dependence of electronic specific heat ($C_{el}/T$) for (Fe,Co,Ni,Cu,Ga)Zr$_2$.

## 4. Conclusion

We synthesized a HEA-type transition-metal zirconide superconductor (Fe,Co,Ni,Cu,Ga)Zr$_2$, which has been designed with 3d transition metals for the *Tr* site. Neutron powder diffraction confirmed the tetragonal CuAl$_2$-type crystal structure and the absence of long-range magnetic ordering in (Fe,Co,Ni,Cu,Ga)Zr$_2$. Although the *a*-axis exhibits normal positive thermal expansion, the lattice constant *c* does not remarkably change at $T$ = 2–290 K, which would be related to the recently-discovered anomalous axes thermal expansion in CoZr$_2$. From electrical resistivity, magnetic susceptibility, and specific heat, the emergence of bulk superconductivity with a $T_c$ of 2.9 K was confirmed. Comparing the experimental data for HEA-type (Fe,Co,Ni,Cu,Ga)Zr$_2$ and zero-entropy CoZr$_2$, we found the temperature dependence of $U_{iso}$s for (Fe,Co,Ni,Cu,Ga)Zr$_2$ are clearly larger than those for CoZr$_2$ at low temperatures. The difference would be caused by the unique phonon characteristics induced by the local disorder in the HEA-type sample. Evidently, the temperature dependence of electrical resistivity exhibits clear difference: large *RRR* for CoZr$_2$ and small *RRR*~1 for the HEA-type sample. The (Fe,Co,Ni,Cu,Ga)Zr$_2$ superconductor would be a model system useful to study the effects of local disorder introduced by the HEA-type site on superconducting properties and to explore exotic superconducting states.

## Acknowledgements

The authors thank M. Fujita, Y. Goto, and O. Miura for their supports in experiments and discussion. This work was performed under the GIMRT Program of the Institute for Materials Research, Tohoku University (CN: Center of Neutron Science for Advanced Materials: Proposal No. 202112-CNKXX-0001). This work was carried out by the JRR-3 program managed by the Institute for Solid State Physics, the University of Tokyo (the T1-3 HERMES IRT program: Proposal No. 22410). This work was partly supported by Grant-in-Aid for Scientific Research (KAKENHI) (Proposal Nos. 21K18834, 21H00151, 19H05164, and 21H00139) and Tokyo Government Advanced Research (H31-1).